\def\year{2020}\relax
\documentclass[letterpaper]{article} 
\usepackage{aaai20}  
\usepackage{times}  
\usepackage{helvet} 
\usepackage{courier}  
\usepackage[hyphens]{url}  
\usepackage{graphicx} 
\usepackage{wrapfig}

\usepackage{color, colortbl}
\usepackage{wrapfig}
\definecolor{Gray}{gray}{0.9}
\definecolor{LightCyan}{rgb}{0.87,0.95,.99}
\definecolor{LightGreen}{rgb}{.87,.99,.87}
\definecolor{LightOrange}{rgb}{.98,.96,.87}
\definecolor{LightPink}{rgb}{.96,.96,.96}

\usepackage{booktabs} 
\usepackage{graphicx}
\usepackage{ wasysym }
\usepackage{graphicx}  
\usepackage{multirow}
\title{The Challenges of Crowd Workers in Rural and Urban America}

\author{Claudia Flores-Saviaga\textsuperscript{\rm 1} Yuwen Li\textsuperscript{\rm 2}, Benjamin V. Hanrahan\textsuperscript{\rm 3},\\ \Large \textbf{Jeffrey Bigham\textsuperscript{\rm 4}, Saiph Savage\textsuperscript{\rm 1,5}}\\ 
\textsuperscript{\rm 1}West Virginia University,
\textsuperscript{\rm 2}University of Washington,
\textsuperscript{\rm 3}Penn State University,\\
\textsuperscript{\rm 4}Carnegie Mellon University,
\textsuperscript{\rm 5}National Autonomous University of Mexico (UNAM)

}
\begin{document}

\maketitle

\begin{abstract}
Crowd work has the potential of helping the financial recovery of regions traditionally plagued by a lack of economic opportunities, e.g., rural areas. However, we currently have limited information about the challenges facing crowd workers from rural and super rural areas as they struggle to make a living through crowd work sites. This paper examines the challenges and advantages of rural and super rural Amazon Mechanical Turk (MTurk) crowd workers and contrasts with those of workers from urban areas. Based on a survey of 421 crowd workers from differing geographic regions in the U.S., we identified how across regions people struggled with being onboarded into crowd work. We uncovered that despite the inequalities and barriers, super rural workers reported earning the highest hourly wages. We also identified cultural traits, relating to time dimension and individualism, that offer us an insight into crowd workers and the necessary qualities for them to succeed on gig platforms. We finish by providing design implications based on our findings to create more inclusive crowd work platforms and tools.

\end{abstract}

\section{Introduction}
The future of work includes new opportunities through online work and the gig economy \cite{graham2017towards,lustig2020stuck}. Crowd work has become a core part of the gig economy, especially because it is an important entry for getting people involved in online labor \cite{idowu2019bright}. Given its flexibility in including non-experts, and because its micro work is not tied to any specific geographic region, crowd work platforms have been named one of the solutions for facilitating the financial recovery of regions traditionally plagued with a lack of economic opportunities, e.g., rural areas \cite{braesemann2020icts}. 

However, for the most part, these crowd work platforms have failed at empowering all geographic regions to access similar economic opportunities \cite{braesemann2020icts}. While crowd work platforms have not been designed to support rural workers, who could benefit the most from crowd work \cite{newlands2020crowdwork}.


While there has been previous research that has focused on the demographics of the gig economy as a whole \cite{broughton2018experiences}, research into the demographics of specifically crowd work (e.g. MTurk) is lacking, especially when it pertains to rural and super rural areas in the U.S. This is important for the simple fact that crowd work is a valuable means to introduce people, especially in remote, underemployed areas, into the gig economy. 
Recently though, we have seen the emergence of new studies that aim to comprehend crowd worker demographics. However, these studies have focused on understanding broadly how crowd workers operate \cite{difallah2018demographics}, their general experiences \cite{martin2014being}, wages \cite{hara2019worker}, and frictions they encounter \cite{mcinnis2016taking,salehi2015we}. Most of these crowd worker studies have rarely considered local U.S. geography into their investigations. The few research that does integrate geography has primarily studied crowd work on specialized knowledge platforms \cite{braesemann2020icts}. Most research has focused on making cross-country comparisons of crowd worker demographics \cite{posch2018characterizing} (e.g., the differences between crowd workers in the U.S. and India \cite{newlands2020crowdwork,martin2016turking,gupta2014turk}). Related work has also focused on freelancing in particular developing countries, but there is not enough work that focuses on crowd work specifically for rural and super rural areas\footnote{https://fair.work/publications/}.

To address this gap, we conducted a survey study contrasting the challenges and advantages of urban, rural, and super rural crowd workers in the U.S. We focus in particular on crowd workers from Amazon Mechanical Turk (MTurk), one of the most popular crowd work platforms. We identified the following themes concerning their experiences: Onboarding, Income, Infrastructure, and Flexibility. Within these themes, we find that super rural workers have the highest hourly wages, despite facing the greatest barriers in infrastructure and education. We also identified essential cultural traits, relating to time and individualism, that offer us an insight into crowd workers and the necessary qualities for them to succeed on gig platforms. Based on our findings, we also discuss design implications which can help to create more inclusive crowd work platforms and tools in order to increase the likelihood that rural and super rural workers utilize crowd platforms as easy and flexible, much-needed job opportunities. As we enter the new post-COVID-19 era, understanding the challenges and advantages of crowd workers from all geographic areas is an essential task to ensure that workers who struggle with unemployment or involuntary part-time employment can have access to the knowledge and tools that could help make them successful workers in the coming work environment.


\section{Related Work}
\label{related}
Researchers have pointed out that crowd work creates opportunities for income and social mobility in regions where local economies may be stagnant \cite{kittur2013future}. This type of digital labour has attracted interest as it can provide economic development opportunities among traditionally excluded populations \cite{newlands2020crowdwork}. Crowd workers have even expressed that they believe that their work on such platforms can contribute in their career advancement \cite{kasunic2019crowd}, while governments consider crowd work to be an alternative route for employment that could extend employment opportunities beyond national geography and regional barriers \cite{idowu2019bright}. However, information is lacking that can differentiate U.S. crowd workers' experiences in rural and super rural areas from their counterparts in urban areas. 

Extant research has started to understand the struggles of crowd workers in developing countries, for example the lack of access to adequate equipment (e.g mobile vs PC) \cite{newlands2020crowdwork}, lack of technical skills, and lack of technical infrastructure in their cities along with the  awareness of the insufficiency of work on the platforms \cite{koskinen2019digital,heeks2017decent}. 
Additionally, existing research has examined how crowd work benefits and challenges specific populations, such as older adults \cite{brewer2016would}, people with disabilities \cite{ding2017socially,zyskowski2015accessible}, people with autism \cite{hara2017introducing}, and people in marginalised regions of the world \cite{wood2018workers}.



Emphasizing the constraints of rural workers is especially important at this point in time as historically these regions have suffered from geographic disparity in terms of economic and social factors. Further research on poverty in rural areas helps define the social and economic struggles faced by these workers \cite{lichter1989underemployment,lichter2002rural,lichter2016people,thiedeLichter2018working} along with statistical projections of continued global growth of rural populations \cite{ILOruralurban2018}. An additional factor to consider is recent research that has exposed the current scarcity of online jobs, which poses an extra hardship for those seeking employment. This is especially the case during the COVID-19 pandemic as more people are creating profiles and seeking freelance work online \cite{stephany2020distancing}.


Previous research has demonstrated that the demographics of MTurk have shifted over time. Before 2008, the majority of the workers were from the United States (76\%) and India (8.03\%), while the rest were from U.K. and Canada, and mostly female (57\%) and young (21-30 years old) \cite{ipeirotis2008mechanical}. In 2010, due to platform changes in payment policies \cite{AmazonMe16:online}, demographics shifted to comprise 57\% of workers from the U.S. and 32\% from India, with a slight decrease in female participation (55\% of participants), and an increase in young population (62\% of participants were between 18-30 years old) \cite{ross2009turkers}. Most recent work shows that most of the workers are from the U.S. (75\%), and India (16\%), while the rest are from Canada, Great Britain, Philippines and Germany; workers appear to be more gender balanced in general (51\% female  and 49\% male); nonetheless, researchers detected significant deviations from the average across countries \cite{difallah2018demographics}. 

Previous work has started to understand rural crowd workers in Europe \cite{vasantha2014social}, finding that flexible hours of working, extra income, and work-life balance are some of the factors that motivate rural workers in Scotland to participate in crowd work. Investigations that contrast urban and rural platform workers from the U.S. have been more limited in studying specialized knowledge platforms, e.g. exploring how many of the most remote regions of the U.S. do not participate in the online labour market at all \cite{braesemann2020icts}. However, previous work has lacked an understanding of super rural crowd workers in the U.S. that do participate in crowd work marketplaces such as MTurk.

The demographics of the crowd workers are receiving attention, precisely because they may determine whether MTurk is appropriate for specific research projects \cite{sandberg2020development,ortega2019crowdsourcing}. 

To build off of this prior work, we do a quantitative and qualitative analysis of survey responses of U.S. crowd workers to understand ways in which workers with these or other constraints living away from urban areas in the U.S. may struggle to benefit from working on these platforms.


\section{Methods}
\label{Method}
Our goal was to identify the challenges and advantages that crowd workers from different geographical regions face. For this purpose, we created a survey that asked crowd workers about their experience on MTurk, as well as the advantages and challenges that they saw about working on the platform. These answers were based on their own personal circumstances, especially the geographical region from which they were from. 

\subsection{Participants}
Similar to prior research \cite{sannon2019privacy}, we recruited participants for our survey via mailing lists of crowd workers we have collected over the years and via HIT postings on MTurk. 
For MTurk we did not ask people to be from a particular geographical region (e.g., rural areas) to avoid that they might misrepresent themselves in order to qualify for the survey \cite{sharpe2017mturk}. Therefore, we opened up the survey to any adult crowd worker. We increased our chance of recruiting rural and super rural workers, by directly emailing rural crowd workers we knew and using snowball sampling to recruit more individuals. We used the diverse voices that participated in the survey to contrast the challenges  of crowd workers from various geographic regions.


Using zip code classifications, we classified workers' geographic locations with assignments of urban, rural, and super rural based on federal standards for remoteness of location \cite{2010Urba6:online,CMSZipCode}.  The Census defines urban as an urbanized area of 50,000 or more people while an urban cluster has a minimum of 2,500 people and a maximum of 50,000 people. While the rural definition encompasses any area that is not included in the urbanized area or urban clusters. Our research focused on how more remote workers engage in crowd work \cite{2010Urba6:online,ratcliffe2016defining}. Thus, we utilized the ``super rural'' designation. The ``super rural'' designation is found in federal designations of ambulance services for extreme remote locations by the Centers for Medicare and Medicaid Services (CMS) \cite{CMSZipCode}. The ``super rural'' designation is achieved by excluding urban zip codes and ranking the remaining zip codes by population density. ``The lowest quartile of these zip code areas are deemed super rural'' while the remaining ``top three quartile zip code areas'' are categorized as rural \cite{Ruralcodes2015}.

\subsection{Data Collection}
\label{data}
Participants were asked to complete a multiple choice and open question survey. The survey elicited information about: 

\begin{itemize}
  \item Workers' General Demographics and Background:
  \begin{itemize}
  \item Zip code, size of the population from where they live. 
  \item Gender, education, and other occupations (aside from MTurk). 
  \end{itemize}
  \item Gig Market Experience:
  \begin{itemize}
  \item How they first heard about MTurk and when they joined the platform.
  \item Tools, forums, and other resources they use for MTurk.
  \item Their hourly wage on MTurk.
  \item Gig markets they are familiar with and gig markets in which they have worked. 
  \item Challenges and advantages they and others from their region face on MTurk.
  \end{itemize}
  \item Cultural Background:
  \begin{itemize}
  \item How they manage and view time. 
  \item Preferences for working on their own or collectively. 
  \end{itemize}
  
\end{itemize}

We modeled many of our demographic and gig market experience questions on previous crowd work research in order to allow for comparison with prior studies \cite{berg2015income,difallah2018demographics}. We also included questions to measure cultural differences and further understand variations between our participants. In particular, we asked workers questions from the value survey model from Hofstede \cite{hofstede2013vsm} and about workers' habits and preferences regarding how they managed their time \cite{oddou1999managing} (which are traditional methods to study cultural variations). Through this, we quantified how much workers self-identified with cultures that: (a) focus on the individual or the collective (this is typically referred to as the individualism dimension, and is measured on a scale from 0-100); (b) emphasize promptness or not, have a time focus to the work they do instead of a relationship focus (this is the time dimension and is measured in a scale from 0-25 \cite{oddou1999managing}). Given that we surveyed crowd workers throughout 2020, we also added questions about how the COVID-19 pandemic had affected their labor on MTurk. After iterating and pilot-testing our (IRB-approved) survey, we had our final survey with 56 questions. Our survey took 10 to 15 minutes to complete. We paid participants \$10 for their participation. 

\begin{figure*}
  \begin{center}
    \includegraphics[width=1\linewidth]{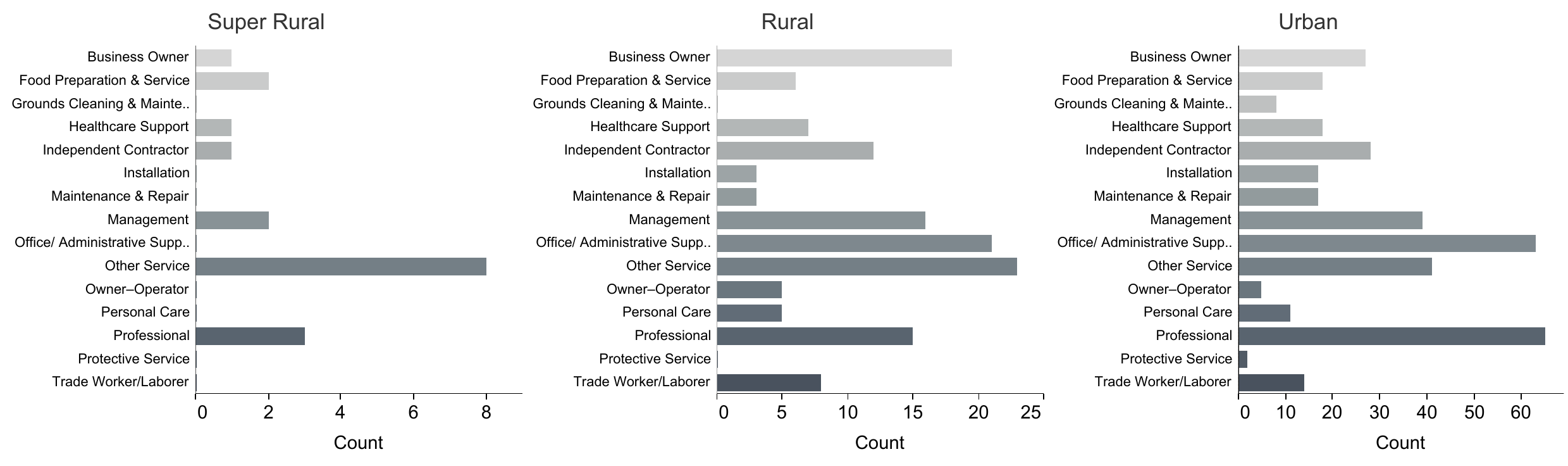}
  \end{center}
    \vspace{-7pt}
  \caption{\bf{Jobs crowd workers had outside MTurk.}}
  \label{fig:Jobs}
  \vspace{-7pt}
\end{figure*}

\subsection{Data Analysis}
We conducted data analysis over the open-ended survey responses of our participants. For our data analysis, we looked for general patterns on crowd workers' responses that summarized their experiences, and highlighted the challenges and advantages that they saw for their region. We aggregated all the open-ended survey responses from participants, as well as all our notes and memos from the study, to begin identifying key concepts and ideas. 

We used open coding to extract initial concepts from the survey \cite{mihas2019qualitative}. We aimed for these initial concepts to take into account some of the themes that related work had derived \cite{posch2018characterizing,kaplan2018striving,martin2014being}. Next, we discussed these initial concepts in their entirety to underscore their importance. With this initial list of codes established, two of the authors then independently coded the data bottom-up and created a set of 11 axial codes which were applied top-down to the survey responses. From the 11 axial codes, we collectively derived a list of four themes representing the different experiences, and general insights that participants reported. Our analysis showed a strong inter-coder agreement (Cohen's Kappa coefficient(k) = 0.826). Disagreements were discussed during the writing and synthesis process. We used our thematic analysis to structure the responses of our participants and highlight the differences and similarities in their experiences. We discovered the following themes representing the different general experiences that crowd workers in our study reported: 

\begin{itemize}
    \item \textbf{Onboarding}: This topic is about the challenges and advantages that exist for integrating new workers onto MTurk, so they can make a living.
    
\item \textbf{Income}: This category relates to the challenges and advantages that exist around the money that is received from crowd work.

\item \textbf{Infrastructure}: This category is about the challenges and advantages that exist around the physical structures and facilities needed to do crowd work.

\item \textbf{Flexibility}: This category is about the challenges and advantages that arise from the ``flexibility'' that crowd work provides. 
\end{itemize}

\section{Results}
\begin{figure}
  \begin{center}
    \includegraphics[width=1.0\linewidth]{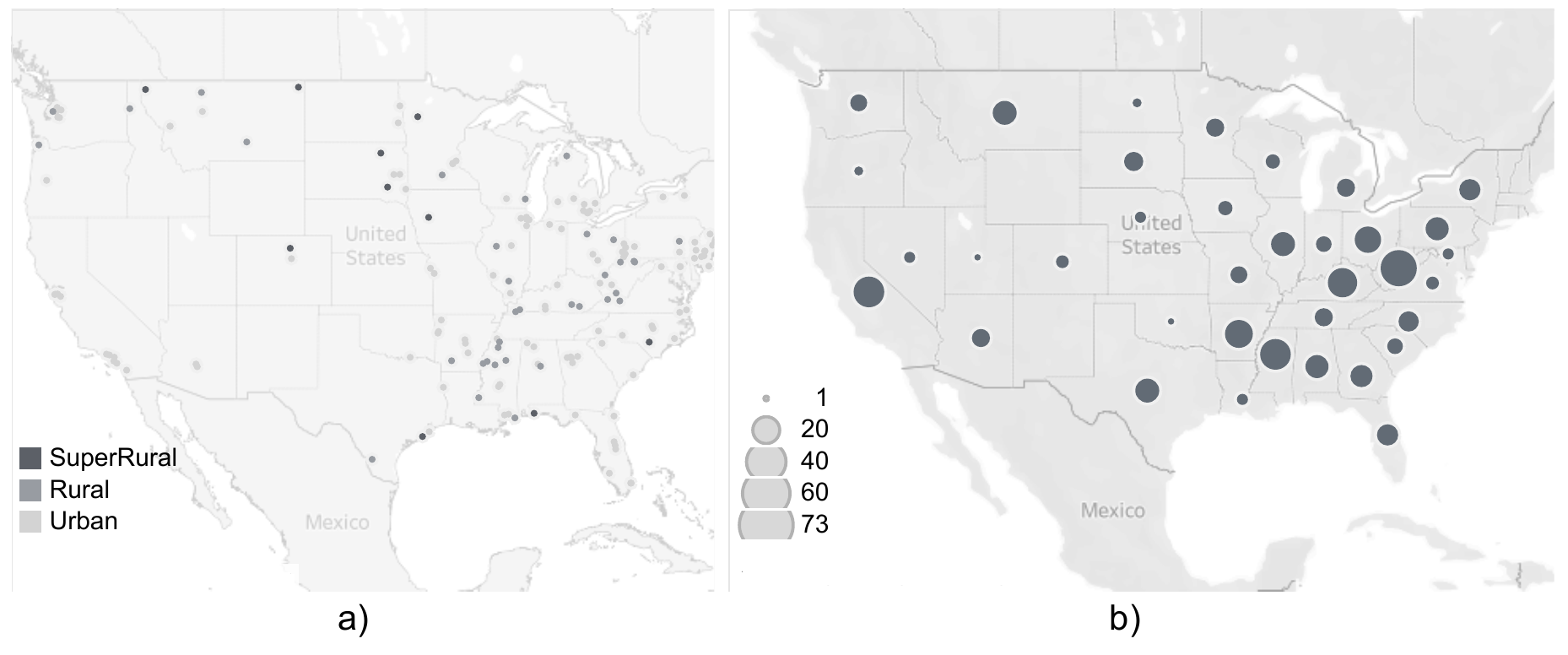}
  \end{center}
  \caption{\bf{Regions where crowd workers came from: (a) shows dots in the regions where workers were from, dots are color coded based on whether they were in super rural, rural, or urban areas; (b) shows the number of workers who reported being from a particular state.}}
  \vspace{-7pt}
  \label{fig:map}
  \vspace{-7pt}
\end{figure}

We had 421 crowd workers who stated in the survey that they lived in the United States: 290 (69\%) lived in urban areas, 114 (27\%) in rural areas, while 17 (4\%) in super rural regions. Fig. \ref{fig:map} shows on a map an overview of the U.S. regions from which our participants reported living. Notice that the groups we study have disproportionate sizes. This is normal when studying groups that make up only a small share of the population \cite{zahnd2019challenges}. We tried to mitigate this with our sampling method where we aimed to incorporate a higher representation of rural and super rural workers; a common practice when studying minority groups to remove the potential for bias \cite{mercer2016oversampling}. 

\subsubsection{Crowd Workers' Age and Gender.}
Our urban and rural survey participants had an average age of 38 (SD=6.36). These demographics are similar to previous survey samples of crowd workers \cite{lascau2019monotasking,kasunic2019crowd}.  
However, super rural workers differed; they were slightly older with an average age of 45 (SD=6.36). With respect to gender, we note that related studies have had fairly even gender splits \cite{berg2015income}. Our data however showed different proportions (especially when comparing urban to rural and super rural). We observed that urban had a larger percentage of men than rural and super rural areas with: 58\% male, 41\% female, and 1\% non-binary. Meanwhile, super rural had the largest percentage of women with: 56\% female, and 44\% male. Rural areas had slightly more men than women, but still less than urban regions with: 52\% male, 47\% female, and 1\% non-binary. Notice that these results match research in rural populations that has observed that rural areas tend to be made up of more women than their urban counterparts, as men migrate to cities to find work \cite{downey1987personal}.

\subsubsection{Crowd Workers' Education.} 
We observed that both urban and rural workers had a relatively high education level, as 68\% of the urban and 63\% of the rural workers reported having completed college or beyond. We observed that for super rural workers this percentage was lower, with only 44\% reporting to have college or beyond.

\subsubsection{Crowd Workers' Occupations Outside MTurk.} 
Across regions, all of our participants had at least one formal job outside MTurk. MTurk did not appear to be the only source of income on which crowd workers relied. It was interesting to observe that the type of jobs that crowd workers had outside MTurk varied across regions. Figure \ref{fig:Jobs} presents a break down of the type of jobs that crowd workers reported having. The most popular external jobs for urban crowd workers were in the categories of ``professional jobs'' and in ``office and administrative support,'' which involved primarily being managers or accountants. For rural and super rural, the most popular job category that they had outside MTurk included ``providing other services''. Upon closer inspection, we noted that this involved primarily supplying their community with different services, such as being a ``neighborhood pet groomer'', ``mowing  neighbors' lawns'', or ``painting houses.'' Rural and super rural workers likely kept a wide range of ``odd jobs'' to cover community needs and earn a living within their community. This matches research that has previously described the female rural workforce in terms of them being a ``discouraged worker \cite{lichter1989underemployment}'' as they struggle with involuntarily working part-time along with a thriving rural ``underground economy'' which includes odd jobs \cite{lichter2016people}. In fact, Lichter et al. describes rural workers as suffering ``less from having no jobs than from having jobs that pay poorly \cite{lichter2002rural}.''


\begin{figure*}
  \begin{center}
    \includegraphics[width=1.0\linewidth]{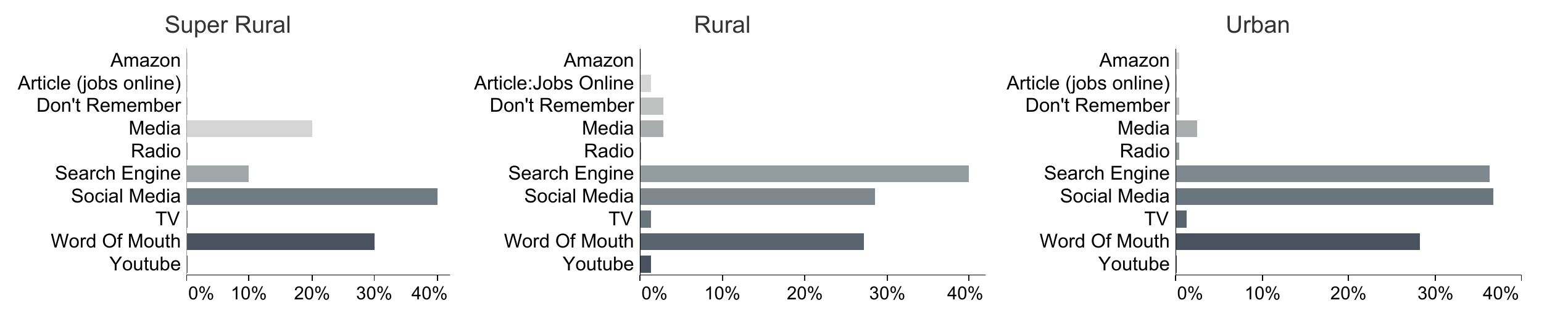}
  \end{center}
  
  \caption{\bf{Sources that first exposed workers to MTurk.}}
  \label{fig:Mturk}
  \vspace{-4pt}
\end{figure*}

\subsubsection{Crowd Workers' Tenure.} We also observed that our urban participants joined MTurk before the rural and super rural populations. The median year for starting to work on MTurk was 2015 for urban workers, whereas it was 2018 for rural and super rural workers. 

\subsubsection{Crowd Workers' Income.}
The reported median hourly wage that urban workers reported was surprisingly lower than what super rural workers reported. Super rural workers were the ones who reported receiving the largest median hourly wage of \$10 an hour, followed by urban workers who reported a median of \$8, and then rural workers with \$7. This finding is interesting as  previous work has reported that only 4\% of crowd workers on MTurk earned more than \$7.25/h \cite{hara2018data}. This might indicate that workers who took the survey had certain qualifications that help them earn higher wages. 


\subsubsection{Exposure to MTurk.}
Figure \ref{fig:Mturk} presents an overview of the different types of sources from which crowd workers reported first hearing about MTurk. We note that workers from super rural regions reported being exposed to MTurk from less varied sources than their urban and rural counterparts. Super rural workers reported only hearing about MTurk from four main sources, whereas rural and urban workers reported over seven different sources that exposed them to MTurk. It was also interesting to observe how the sources that were the most popular for exposing people to MTurk varied across regions. For urban workers, their main source for hearing about MTurk was social media (37\%), while search engines was second (36\%). Whereas for rural individuals their main source was search engines (40\%), while social media was second (29\%). However, for super rural workers, their main source was social media (40\%), while word of mouth was second (30\%). This matches research that has shown that word of mouth is an effective source of recruitment in rural regions ``partly due to its credibility \cite{van2016social}.'' It also highlights the economic value of credibility within super rural areas as novel information is shared \cite{granovetter2005impact}.

\subsubsection{Crowd Workers' Learning Resources.}
We also asked participants which tools (if any) they used to improve on MTurk (e.g., to develop their skills) and thus increase their earnings. We observed that, in general, urban workers (81\%) made more use of online forums and tools, such as Turker View or Turkopticon, to grow on MTurk than what the rural (73\%) and super rural (69\% ) population reported using.


\subsubsection{Crowd Workers and Other Gig Platforms.}
Of our total crowd workers, only 43\% had worked at least once on a gig platform aside from MTurk. When we break down this statistic into regions, we see that the super rural workers (30\%) were the ones who least frequently worked on other gig platforms in comparison to urban (34\%) and rural workers (36\%). This finding is not unexpected as work on many gig platforms is often location-based. To better understand this finding, we also investigated how frequently the workers in each region performed location-based gig work and non-location-based gig work (i.e., work on gig markets that is not dependent on location). We found that our rural (23\%) and super rural (29\%) workers were slightly more present on non-location-based gig platforms (e.g., Upwork or Etsy) than urban workers (21\%); while across regions, our participants rarely worked on gig markets that were location-based, with only 13\% of urban and rural workers participating and super rural working even less often with 9\%. (Note that our location and non-location-based statistics are not mutually exclusive.)

\subsubsection{Crowd Workers' Voice: Challenges and Advantages.}
\label{surveyhabits}
\begin{figure}
  \begin{center}
    \includegraphics[width=1.0\linewidth]{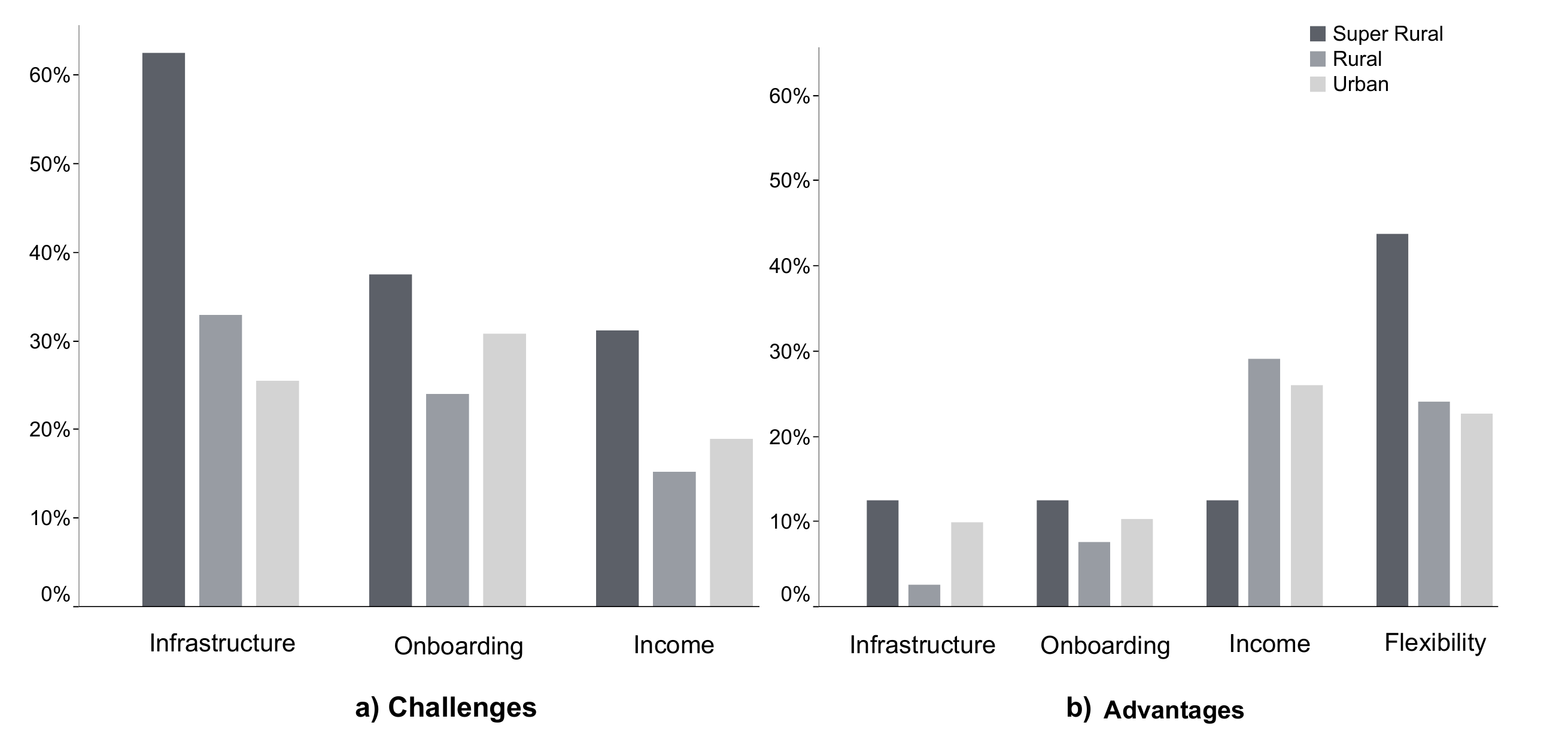}
  \end{center}
  \caption{\bf{Overview of the challenges and advantages that crowd workers identified.}}
  \vspace{-7pt}
  \label{fig:challenges}
  \vspace{-7pt}
\end{figure}

 Figure \ref{fig:challenges} presents an overview of the percentage of workers who discussed experiencing certain challenges (Fig \ref{fig:challenges}a.) and advantages (Fig \ref{fig:challenges}b.) when performing crowd work in their region. The categories come from our thematic analysis. \newline
 \newline
 
\textbf{Infrastructure}.

 One of the main challenges that crowd workers discussed across regions was infrastructure, with super rural workers (63\%) stating the most that infrastructure posed a challenge for them compared to rural (33\%) and urban (26\%) workers. For rural and super rural workers, the main problem associated with infrastructure was having access to high-speed broadband. As one super rural worker mentioned:
 
\begin{quote}
    \textit{``
...My town is quite rural and they barely got decent internet; so it's pretty much an uphill battle for most of them.'' SR\_66}.
\end{quote}

Urban workers discussed challenges associated with infrastructure to a lesser degree than rural and super rural workers, while also discussing this challenge in a different way. For urban workers, the main challenges related to infrastructure were about deciding to work in different parts of the city, and suddenly encountering ``spotty internet'':

\begin{quote}
    \textit{``Lots of working class people here can use it [MTurk] on the go to work, while out and about or on breaks at work [...] but of course they might have spotty internet connectivity...'' U\_247}.
\end{quote}

This notion of being able to use a city's infrastructure to work from anywhere was also seen as an advantage for these workers:

 \begin{quote}
    \textit{``I live in a city so, wifi is most everywhere. If one doesn't have internet access at home, there are a lot of places (restaurants, libraries, etc) that have it available for free.'' U1}. 
\end{quote}

Our participants across regions rarely discussed advantages related to infrastructure and MTurk, with rural workers (3\%) discussing the advantages much less than urban (10\%) workers. However, it is important to note that super rural (13\%) workers, the ones who discussed the challenges of infrastructure the most, also discussed the advantages related to infrastructure the most. Some workers from these remote areas felt that MTurk could force individuals to have better infrastructure (e.g., better internet connectivity):


\begin{quote}
    \textit{``[the advantages are that] in many cases MTurk makes us have faster Internet. Better home situation...'' SR\_{354}}.
\end{quote}

\textbf{Onboarding process}.

Other challenges discussed across regions were the ``onboarding process'' and the ``low wages'' on the platform. The super rural workers (38\%) were the ones that discussed the challenges with onboarding the most, while urban workers (31\%) and rural workers (24\%) discussed these challenges less. The onboarding process proved challenging due to all the learning that novice workers had to do in order to identify legitimate labor and start making wages (e.g., they had to learn what tools to use or how to screen HITs).


\begin{quote}
    \textit{``[the biggest challenge is to] learn the scripts, learn to look up reviews, learn that it's ok to turn stuff down if it pays pennies.'' U\_145}.
\end{quote} 

Workers also discussed how MTurk did not facilitate the onboarding process, especially because the platform put limitations into the type of labor that novices can access:

\begin{quote}
    \textit{``...new workers are limited in how many HITs they can perform each day and they often lack access to higher paying HITs. This frustrates new people...'' U\_134}.
\end{quote} 

Our participants across regions expressed that one of the things that would make MTurk easier for people from their region was to have better onboarding processes:

\begin{quote}
    \textit{``Flatten the learning curve by actually teaching MTurk workers the basics of what they need to know and improve the overall system.'' R\_372}.
\end{quote}

\begin{quote}
    \textit{``Easier learning curve for people just starting out, better resources for explaining how to maximize output'' U\_98}.
\end{quote}

It was interesting to see however that workers also saw advantages to having a difficult onboarding process. Similar to infrastructure, super rural workers (13\%) also discussed the advantages of onboarding the most, compared with urban (10\%) and rural workers (8\%). Super rural workers discussed how such a type of onboarding process could help people in their village to develop digital skills:

\begin{quote}
    \textit{``[the advantages are that] people here [in the super rural area] have the time to learn the things needed to be able to get decent HITs and it could help them have new computer skills...'' SR\_143}.
\end{quote}

Urban and rural populations also saw advantages to the difficulties of the onboarding process. In particular, they believed their region had several individuals who were significantly tech savvy, and as a result it would enable them to access jobs that not everyone in the U.S. could do (as the onboarding process would be a barrier): 

\begin{quote}
    \textit{``[the advantages are that it's a] heavy college town. They [people living in her rural area] are technologically adept and able to work on MTurk. So this would give them [people living in her rural area] an advantage...'' R\_218}.
\end{quote}

\begin{quote}
    \textit{``[the advantages are that] my area has a well educated populace so that would make MTurk tasks easier to understand for area workers and the requesters would have more quality work from residents in my area'' U\_38}.
\end{quote}

\textbf{Income}.

The other main challenge that workers across geographical regions identified was the income they earned on MTurk, with super rural workers (31\%) discussing income the most followed by urban (19\%) and rural (15\%) workers. Although, hardships around income were discussed slightly differently in rural and urban regions. For rural and super rural workers, the main challenge associated with income was that it was not constant and steady:

\begin{quote}
    \textit{``[the main challenge] is not having enough tasks available to get a steady paycheck...'' R\_248}.
\end{quote}

For urban workers however the challenge was that the income was too low to cover their costs of living. This meant they needed to invest long hours on the platform to make ends meet:

\begin{quote}
    \textit{``[The challenge is] the amount of work needed to be done in order to make ends meet is difficult to perform due to the high cost of living in my area.'' U38}.
\end{quote}


It was interesting to see that for rural and super rural workers, unlike their urban counterparts, the compensation earned from MTurk was seen as an advantage. Rural participants rarely discussed problems with making ends meet. Instead, they discussed how MTurk could help address the problem of limited job opportunities in their regions:

\begin{quote}
    \textit{``...A lot of people [in my rural town] only have part time jobs or none at all [...] There are not many jobs where I am from and it is very remote geographically speaking [...] The extra work [from MTurk] would benefit my community and their families by helping to provide food and shelter..'' R\_{285}}. 
\end{quote}

Additionally, the income received was seen as worth it by rural and super rural workers (likely because the lack of economic development in rural regions can result in ``intergenerational immobility \cite{gornick2016gender}'' and chronic poverty \cite{lichter2016people}):

 \begin{quote}
    \textit{``I think people from my hometown are probably more used to doing low-paid work, since it's a small town and wages are low here. That might lead to people being more willing to stick with the platform [MTurk] to earn money over time..'' R51}. 
\end{quote}

For urban individuals, the advantages that MTurk income brought was that it helped them to complement the income they received from their full-time jobs:

\begin{quote}
    \textit{``[the advantages are that it's] possible to hold a full time job and turk at the same time [...] It's extra money I can make in spare time in order to supplement income and help pay the bills...'' U\_{82}}. 
\end{quote}
\textbf{Flexibility.}

In terms of benefits, one of the main advantages that super rural (44\%) and rural (24\%) workers identified with crowd work was the flexibility to work from home and not use transportation:

\begin{quote}
    \textit{``[the advantages are that this town] is very rural and it would be an easy way to make money from home.'' SR\_{342}}. 
\end{quote}

\begin{quote}
    \textit{``[The main advantage is that it is] simple work, able to be done from home without driving dangerous roads.'' R\_{245}}. 
\end{quote}


Urban workers (23\%) also discussed the flexibility of crowd work as an advantage. 
However, they appreciated it because the work provided them with the flexibility to have multiple jobs: 

\begin{quote}
    \textit{``[The main advantage is that it's] possible to hold a full time job and turk at the same time.'' U\_{351}}. 
\end{quote}

\subsection{Crowd Workers and Culture.}
We calculated the median ``culture scores'' of workers per region for the cultural dimensions of ``individualism'' and ``time''. Across regions, crowd workers had surprisingly similar culture scores. The median individualism score was 65 and the median time score was 12. Results of an independent-samples Mann-Whitney test indicated that there was no statistical difference among groups in their culture scores (p-value=0.92 for individualism index; p-value= 0.02 for time). This is noteworthy given that prior work has reported possible cultural differences between rural and urban areas with terms such as ``cultural isolation''   \cite{lichter2016people} and ``cultural balkanization'' \cite{lichter2002rural,lichter2011rural}. Our results however might hint that regardless of whether they are rural or urban, crowd workers on MTurk share similar cultural characteristics. In our case, our crowd workers showcased a high individualism score, which is typical of the U.S. culture \cite{caldwell2006personality}. A high individualism score means they value the performance of individuals over groups. The low time score indicates that workers in our study tended to emphasize promptness, had a short term perspective, and tended to be more focused on tasks and job completion than in maintaining relationships \cite{hofstede1984cultural}.

Finally, we asked crowd workers about whether they experienced any interruptions due to COVID-19. The majority (27.4\%) stated they were working as normal while 9.26\% expressed having had some interruptions. Specifically, they stated seeing more HITs related to COVID-19, and some (especially urban workers) had to work on MTurk from home instead of in their free time at their offices.

\section{Discussion}
\label{Disc}
In this research, we study the challenges and opportunities that urban, rural and super rural crowd workers experience on MTurk. In this section, we present our discussion based on our quantitative and qualitative data from our survey responses, and we also connect our results with prior literature. 

\subsubsection{Super Rural America \& MTurk Wage Opportunities.}
Our survey highlighted that super rural workers had slightly more barriers than their urban counterparts. For example, they joined MTurk later, had less education, heard from MTurk from a more limited number of sources, used less tools than urban workers, and tended to report more struggles with their infrastructure for working on MTurk. However, despite these barriers, super rural workers were the ones who somehow managed to gain the highest median hourly wage. Previous research has identified that when crowd workers believe they are being paid fairly, they tend to have better work results on the platform \cite{ye2017does,deng2016microtask}. In our survey, super rural workers in general did not mention challenges associated with earning a fair wage on MTurk. The main challenges they identified concerning payments revolved more around being able to make a frequent paycheck. It is likely that because super rural workers felt in general that they were receiving fair wages, they stuck more with the platform and eventually earned more money than their urban counterparts. Lichter et al. \cite{lichter2016people} has described that super rural workers are known by their ``self-reliance, independence, and hard work.'' All of this together likely helped super rural workers to strive more on MTurk. Notice also that our results connect to very recent research comparing urban and rural gig workers \cite{braesemann2020icts}. The work found that rural workers tended to be more skilled at gig work than their urban counterparts. While the research did not study crowd work, it is interesting that our results hint that the super rural are also striving more in micro-tasking. 


\subsubsection{Cultural Differences and Crowd Work.}
According to our results, the differences in the cultural dimensions of individualism and time between rural, super rural, and urban crowd workers was not statistically significant. This could be interpreted as the MTurk platform is likely attracting workers with a similar cultural mindset, regardless of the geographical region from where they are from. We argue that analyzing crowd workers' cultural dimensions of time and individualism could offer us an insight into the core cultural traits which might be necessary for striving in micro work. 

Tietze et al. \cite{tietze2002work} has described that some of the traits that are essential for working from home involve being able to use time as ``a central factor in shaping the experience of work,'' and also being able to act as individuals capable of ``becoming the authors of our own lives.'' Our participants with their lower time dimension scores also showed a tendency towards being preoccupied with time to shape and drive the labor they do at home \cite{duranti2009everything}. This preoccupation with time is likely also helpful for crowd work where there is a need to be ``always on call'' and hypervigilant in order to get the higher paying tasks before they are gone \cite{gray2019ghost,williams2019perpetual,lascau2019monotasking}. It is unclear whether our participants have always had this cultural mindset, or whether MTurk contributed to changing how they feel about time. What is important to identify is that workers in our study, based on their time dimension scores, seemed to treat time as a commodity of high value, something that is necessary or perhaps even more important than satisfaction \cite{duranti2009everything}. Nevertheless, it is important to also be aware that the International Labour Organization (ILO) describes how pushing workers to be ``on call'' can also create challenges for many workers by potentially reducing their earning potential and can lead to work-life imbalances \cite{ILOoncall2016}. As previous research has shown that technology can constrain and shape the actions of users \cite{latour1992missing}, we see these findings as critical to understanding how crowd workers thrive on the platform, and also better understanding the types of lives they live outside crowd labor \cite{williams2019perpetual}.

Concerning the individualism dimension: literature has stressed that societies with high scores in this dimension have a high degree of interdependence among its members. In individualist cultures (such as the US \cite{geerthof8:online}), the relationship between the employee and employer tends to be a business relationship based on the assumption of mutual advantage \cite{hofstede1984cultural}. Crowd work in general has focused on providing advantages to requesters \cite{gray2019ghost,mcinnis2016taking}. We believe there is likely value in exploring interfaces that per geographical location can question workers on the advantages they see for working on crowd work in their specific location, and then highlight to others in the region how they can also take advantage.

\subsubsection{Implications for Design.}
Our results showed that the onboarding process was challenging across geographical groups due to the learning curve that workers had to overcome just to start making earnings. We believe there is value for designers to build tools for facilitating workers' onboarding process. These tools could focus on helping workers to develop ``gig literacy skills'' (i.e., the skills needed to start making money within online gig work) \cite{sutherland2020work,suzuki2016atelier}, and also new computer skills that could translate to jobs outside MTurk \cite{kasunic2019crowd}. Our results hint that this might benefit especially super rural crowd workers, who had the less formal education. Crowd work could become an important space that empowers these populations to strive and grow while making a living. It is important however that when we design these tools they focus on not only keeping track of workers' development in the crowd work platform, but also offers workers transferability to other online labor markets, and especially jobs outside MTurk. For instance, it could be beneficial if workers are able to have a way to demonstrate their career/skill advancement when they apply to a job outside of MTurk. This way workers can have the flexibility to keep advancing their career on crowd work platforms, while at the same time having the possibility to prove to outsiders that they have a certain level of competency. 

We also argue that it is important for crowd work platforms to have designs that are better adapted to rural regions, especially as they have the potential of positively affecting the local economy of the region at large, and we saw through our study how engaged rural workers were. This adaptation can involve designing interfaces that are more gender focused (e.g., for rural women).  It could also involve designing intervention programs that allow rural adults to work more from home. While prior work has designed interfaces and interventions for rural areas that facilitate in person mobile gig work \cite{nytimes,newlands2020crowdwork}, our work highlights that US rural adults likely prefer options where they can limit the amount of driving they do. 


\subsubsection{Limitations and Future Work}
\label{limitations} Our participants were active MTurk workers. Future work could focus on studying workers who dropped out of MTurk. We also recruited individuals who were willing to participate in a survey. We are therefore missing the workers who opt out of surveys. We aimed to overcome this limitation by directly mailing workers and personally inviting them to our study. Notice also that we had a varied number of rural, super rural, and urban participants. This disproportion is normal when working with minority groups \cite{zahnd2019challenges}. However, we adapted our recruitment method as best as possible in order to include in our study the voices of workers in rural and super rural areas locations that would otherwise be difficult to reach and document. Further investigations could focus on conducting interviews with participants from these regions to understand in depth why, despite the challenges they face, rural and super rural workers are able to succeed in crowd work (sometimes even more than their urban counterparts). Future research could also study in more depth the cultural background of crowd workers across geographical regions. 


\bibliographystyle{aaai}
\bibliography{sample-base.bib}

\begin{thebibliography}{}

\bibitem[\protect\citeauthoryear{Berg}{2015}]{berg2015income}
Berg, J.
\newblock 2015.
\newblock Income security in the on-demand economy: Findings and policy lessons
  from a survey of crowdworkers.
\newblock {\em Comp. Lab. L. \& Pol'y J.} 37:543.

\bibitem[\protect\citeauthoryear{Braesemann \bgroup et al\mbox.\egroup
  }{2020}]{braesemann2020icts}
Braesemann, F.; Lehdonvirta, V.; and K{\"a}ssi, O.
\newblock 2020.
\newblock Icts and the urban-rural divide: can online labour platforms bridge
  the gap?
\newblock {\em Information, Communication \& Society}  1--21.

\bibitem[\protect\citeauthoryear{Brewer \bgroup et al\mbox.\egroup
  }{2016}]{brewer2016would}
Brewer, R.; Morris, M.~R.; and Piper, A.~M.
\newblock 2016.
\newblock " why would anybody do this?" understanding older adults' motivations
  and challenges in crowd work.
\newblock In {\em Proceedings of the 2016 CHI Conference on Human Factors in
  Computing Systems},  2246--2257.

\bibitem[\protect\citeauthoryear{Broughton \bgroup et al\mbox.\egroup
  }{2018}]{broughton2018experiences}
Broughton, A.; Gloster, R.; and Marvell, R.
\newblock 2018.
\newblock The experiences of individuals in the gig economy. london: Department
  for business.
\newblock {\em Energy and Industrial Strategy}.

\bibitem[\protect\citeauthoryear{Bureau}{2010}]{2010Urba6:online}
Bureau, U.~C.
\newblock 2010.
\newblock Urban area faqs.
\newblock
  \url{https://www.census.gov/programs-surveys/geography/about/faq/2010-urban-area-faq.html}.

\bibitem[\protect\citeauthoryear{Caldwell-Harris \bgroup et al\mbox.\egroup
  }{2006}]{caldwell2006personality}
Caldwell-Harris, C.~L., and Aycicegi, A.
\newblock 2006.
\newblock When personality and culture clash: The psychological distress of
  allocentrics in an individualist culture and idiocentrics in a collectivist
  culture.
\newblock {\em Transcultural psychiatry} 43(3):331--361.

\bibitem[\protect\citeauthoryear{Deng \bgroup et al\mbox.\egroup
  }{2016}]{deng2016microtask}
Deng, X.~N.; Joshi, K.; and Galliers, R.~D.
\newblock 2016.
\newblock Microtask crowdsourcing can both empower and marginalise workers.
\newblock {\em LSE Business Review}.

\bibitem[\protect\citeauthoryear{Difallah \bgroup et al\mbox.\egroup
  }{2018}]{difallah2018demographics}
Difallah, D.; Filatova, E.; and Ipeirotis, P.
\newblock 2018.
\newblock Demographics and dynamics of mechanical turk workers.
\newblock In {\em Proceedings of the eleventh ACM international conference on
  web search and data mining},  135--143.

\bibitem[\protect\citeauthoryear{Ding \bgroup et al\mbox.\egroup
  }{2017}]{ding2017socially}
Ding, X.; Shih, P.~C.; and Gu, N.
\newblock 2017.
\newblock Socially embedded work: A study of wheelchair users performing online
  crowd work in china.
\newblock In {\em Proceedings of the 2017 ACM Conference on Computer Supported
  Cooperative Work and Social Computing},  642--654.

\bibitem[\protect\citeauthoryear{Downey \bgroup et al\mbox.\egroup
  }{1987}]{downey1987personal}
Downey, G., and Moen, P.
\newblock 1987.
\newblock Personal efficacy, income, and family transitions: A longitudinal
  study of women heading households.
\newblock {\em Journal of Health and Social Behavior}  320--333.

\bibitem[\protect\citeauthoryear{Duranti \bgroup et al\mbox.\egroup
  }{2009}]{duranti2009everything}
Duranti, G., and Di~Prata, O.
\newblock 2009.
\newblock {\em Everything is about Time: Does it Have the Same Meaning All Over
  the World?}
\newblock Project Management Institute.

\bibitem[\protect\citeauthoryear{Gornick \bgroup et al\mbox.\egroup
  }{2016}]{gornick2016gender}
Gornick, J.~C., and Boeri, N.
\newblock 2016.
\newblock Gender and poverty.
\newblock {\em The Oxford handbook of the social science of poverty}  221--46.

\bibitem[\protect\citeauthoryear{Gov}{2019}]{CMSZipCode}
Gov, C.
\newblock 2019.
\newblock Ambulance fee schedule.
\newblock
  \url{https://www.cms.gov/Medicare/Medicare-Fee-for-Service-Payment/AmbulanceFeeSchedule}.

\bibitem[\protect\citeauthoryear{Graham}{2017}]{graham2017towards}
Graham, M.
\newblock 2017.
\newblock {\em Towards a fairer gig reconomy}.
\newblock Meatspace Press.

\bibitem[\protect\citeauthoryear{Granovetter}{2005}]{granovetter2005impact}
Granovetter, M.
\newblock 2005.
\newblock The impact of social structure on economic outcomes.
\newblock {\em Journal of economic perspectives} 19(1):33--50.

\bibitem[\protect\citeauthoryear{Gray \bgroup et al\mbox.\egroup
  }{2019}]{gray2019ghost}
Gray, M.~L., and Suri, S.
\newblock 2019.
\newblock {\em Ghost Work: How to Stop Silicon Valley from Building a New
  Global Underclass}.
\newblock Eamon Dolan Books.

\bibitem[\protect\citeauthoryear{Gupta \bgroup et al\mbox.\egroup
  }{2014}]{gupta2014turk}
Gupta, N.; Martin, D.; Hanrahan, B.~V.; and O'Neill, J.
\newblock 2014.
\newblock Turk-life in india.
\newblock In {\em Proceedings of the 18th International Conference on
  Supporting Group Work},  1--11.

\bibitem[\protect\citeauthoryear{Hara \bgroup et al\mbox.\egroup
  }{2017}]{hara2017introducing}
Hara, K., and Bigham, J.~P.
\newblock 2017.
\newblock Introducing people with asd to crowd work.
\newblock In {\em Proceedings of the 19th International ACM SIGACCESS
  Conference on Computers and Accessibility},  42--51.

\bibitem[\protect\citeauthoryear{Hara \bgroup et al\mbox.\egroup
  }{2018}]{hara2018data}
Hara, K.; Adams, A.; Milland, K.; Savage, S.; Callison-Burch, C.; and Bigham,
  J.~P.
\newblock 2018.
\newblock A data-driven analysis of workers' earnings on amazon mechanical
  turk.
\newblock In {\em Proceedings of the 2018 CHI Conference on Human Factors in
  Computing Systems},  1--14.

\bibitem[\protect\citeauthoryear{Hara \bgroup et al\mbox.\egroup
  }{2019}]{hara2019worker}
Hara, K.; Adams, A.; Milland, K.; Savage, S.; Hanrahan, B.~V.; Bigham, J.~P.;
  and Callison-Burch, C.
\newblock 2019.
\newblock Worker demographics and earnings on amazon mechanical turk: An
  exploratory analysis.
\newblock In {\em Extended Abstracts of the 2019 CHI Conference on Human
  Factors in Computing Systems},  1--6.

\bibitem[\protect\citeauthoryear{Heeks}{2017}]{heeks2017decent}
Heeks, R.
\newblock 2017.
\newblock Decent work and the digital gig economy: a developing country
  perspective on employment impacts and standards in online outsourcing,
  crowdwork, etc.
\newblock {\em Development Informatics Working Paper}.

\bibitem[\protect\citeauthoryear{Hofstede \bgroup et al\mbox.\egroup
  }{2013}]{hofstede2013vsm}
Hofstede, G., and Minkov, M.
\newblock 2013.
\newblock Vsm 2013.
\newblock {\em Values survey module}.

\bibitem[\protect\citeauthoryear{Hofstede}{1984}]{hofstede1984cultural}
Hofstede, G.
\newblock 1984.
\newblock Cultural dimensions in management and planning.
\newblock {\em Asia Pacific journal of management} 1(2):81--99.

\bibitem[\protect\citeauthoryear{Hofstede}{2020}]{geerthof8:online}
Hofstede, G.
\newblock 2020.
\newblock geerthofstede.nl.
\newblock \url{http://www.geerthofstede.nl/}.

\bibitem[\protect\citeauthoryear{Idowu \bgroup et al\mbox.\egroup
  }{2019}]{idowu2019bright}
Idowu, A., and Elbanna, A.
\newblock 2019.
\newblock Bright ict and unbounded employment: Typology of crowdworkers and
  their lived and envisaged career trajectory in nigeria.
\newblock In {\em International Working Conference on Transfer and Diffusion of
  IT},  470--486.
\newblock Springer.

\bibitem[\protect\citeauthoryear{ILO}{2016}]{ILOoncall2016}
ILO.
\newblock 2016.
\newblock What are part-time and on-call work?
\newblock
  \url{https://www.ilo.org/global/topics/non-standard-employment/WCMS_534825/lang--en/index.htm}.

\bibitem[\protect\citeauthoryear{ILO}{2018}]{ILOruralurban2018}
ILO.
\newblock 2018.
\newblock Rural-urban labour statistics.
\newblock
  \url{https://www.ilo.org/wcmsp5/groups/public/---dgreports/---stat/documents/meetingdocument/wcms_636038.pdf}.

\bibitem[\protect\citeauthoryear{Ipeirotis}{2008}]{ipeirotis2008mechanical}
Ipeirotis, P.
\newblock 2008.
\newblock Mechanical turk: The demographics. a computer scientist in a business
  school.

\bibitem[\protect\citeauthoryear{Kaplan \bgroup et al\mbox.\egroup
  }{2018}]{kaplan2018striving}
Kaplan, T.; Saito, S.; Hara, K.; and Bigham, J.~P.
\newblock 2018.
\newblock Striving to earn more: a survey of work strategies and tool use among
  crowd workers.
\newblock In {\em Sixth AAAI Conference on Human Computation and
  Crowdsourcing}.

\bibitem[\protect\citeauthoryear{Kasunic \bgroup et al\mbox.\egroup
  }{2018}]{kasunic2019crowd}
Kasunic, A.; Chiang, C.-W.; Kaufman, G.; and Savage, S.
\newblock 2018.
\newblock Crowd work on a cv? understanding how amt fits into turkers' career
  goals and professional profiles.
\newblock {\em Collective Intelligence.}

\bibitem[\protect\citeauthoryear{Kittur \bgroup et al\mbox.\egroup
  }{2013}]{kittur2013future}
Kittur, A.; Nickerson, J.~V.; Bernstein, M.; Gerber, E.; Shaw, A.; Zimmerman,
  J.; Lease, M.; and Horton, J.
\newblock 2013.
\newblock The future of crowd work.
\newblock In {\em Proceedings of the 2013 conference on Computer supported
  cooperative work},  1301--1318.

\bibitem[\protect\citeauthoryear{Koskinen \bgroup et al\mbox.\egroup
  }{2019}]{koskinen2019digital}
Koskinen, K.; Bonina, C.; and Eaton, B.
\newblock 2019.
\newblock Digital platforms in the global south: foundations and research
  agenda.
\newblock In {\em International Conference on Social Implications of Computers
  in Developing Countries},  319--330.
\newblock Springer.

\bibitem[\protect\citeauthoryear{Lascau \bgroup et al\mbox.\egroup
  }{2019}]{lascau2019monotasking}
Lascau, L.; Gould, S.~J.; Cox, A.~L.; Karmannaya, E.; and Brumby, D.~P.
\newblock 2019.
\newblock Monotasking or multitasking: Designing for crowdworkers' preferences.
\newblock In {\em Proceedings of the 2019 CHI Conference on Human Factors in
  Computing Systems},  1--14.

\bibitem[\protect\citeauthoryear{Latour}{1992}]{latour1992missing}
Latour, B.
\newblock 1992.
\newblock Where are the missing masses? the sociology of a few mundane
  artifacts. shaping technology/building society: studies in sociotechnical
  change. we bijker and j. law.
\newblock {\em Bijker and John Law} 225:258.

\bibitem[\protect\citeauthoryear{Lichter \bgroup et al\mbox.\egroup
  }{2002}]{lichter2002rural}
Lichter, D.~T., and Jensen, L.
\newblock 2002.
\newblock Rural america in transition.
\newblock {\em Rural dimensions of welfare reform} ~77.

\bibitem[\protect\citeauthoryear{Lichter \bgroup et al\mbox.\egroup
  }{2011}]{lichter2011rural}
Lichter, D.~T., and Brown, D.~L.
\newblock 2011.
\newblock Rural america in an urban society: Changing spatial and social
  boundaries.
\newblock {\em Annual review of sociology} 37:565--592.

\bibitem[\protect\citeauthoryear{Lichter \bgroup et al\mbox.\egroup
  }{2016}]{lichter2016people}
Lichter, D.~T., and Schafft, K.~A.
\newblock 2016.
\newblock People and places left behind.
\newblock {\em The oxford handbook of the social science of poverty}  317.

\bibitem[\protect\citeauthoryear{Lichter}{1989}]{lichter1989underemployment}
Lichter, D.~T.
\newblock 1989.
\newblock The underemployment of american rural women: Prevalence, trends and
  spatial inequality.
\newblock {\em Journal of Rural Studies} 5(2):199--208.

\bibitem[\protect\citeauthoryear{Lustig \bgroup et al\mbox.\egroup
  }{2020}]{lustig2020stuck}
Lustig, C.; Rintel, S.; Scult, L.; and Suri, S.
\newblock 2020.
\newblock Stuck in the middle with you: The transaction costs of corporate
  employees hiring freelancers.
\newblock {\em Proceedings of the ACM on Human-Computer Interaction}
  4(CSCW1):1--28.

\bibitem[\protect\citeauthoryear{Martin \bgroup et al\mbox.\egroup
  }{2014}]{martin2014being}
Martin, D.; Hanrahan, B.~V.; O'Neill, J.; and Gupta, N.
\newblock 2014.
\newblock Being a turker.
\newblock In {\em Proceedings of the 17th ACM conference on Computer supported
  cooperative work \& social computing},  224--235.

\bibitem[\protect\citeauthoryear{Martin \bgroup et al\mbox.\egroup
  }{2016}]{martin2016turking}
Martin, D.; O’Neill, J.; Gupta, N.; and Hanrahan, B.~V.
\newblock 2016.
\newblock Turking in a global labour market.
\newblock {\em Computer Supported Cooperative Work (CSCW)} 25(1):39--77.

\bibitem[\protect\citeauthoryear{McInnis \bgroup et al\mbox.\egroup
  }{2016}]{mcinnis2016taking}
McInnis, B.; Cosley, D.; Nam, C.; and Leshed, G.
\newblock 2016.
\newblock Taking a hit: Designing around rejection, mistrust, risk, and
  workers' experiences in amazon mechanical turk.
\newblock In {\em Proceedings of the 2016 CHI conference on human factors in
  computing systems},  2271--2282.

\bibitem[\protect\citeauthoryear{Mercer}{2016}]{mercer2016oversampling}
Mercer, A.
\newblock 2016.
\newblock Oversampling is used to study small groups, not bias poll results.

\bibitem[\protect\citeauthoryear{Mihas}{2019}]{mihas2019qualitative}
Mihas, P.
\newblock 2019.
\newblock Qualitative data analysis.
\newblock In {\em Oxford Research Encyclopedia of Education}. Oxford.

\bibitem[\protect\citeauthoryear{Newlands \bgroup et al\mbox.\egroup
  }{2020}]{newlands2020crowdwork}
Newlands, G., and Lutz, C.
\newblock 2020.
\newblock Crowdwork and the mobile underclass: Barriers to participation in
  india and the united states.
\newblock {\em new media \& society}  1461444820901847.

\bibitem[\protect\citeauthoryear{Nosorh}{2015}]{Ruralcodes2015}
Nosorh.
\newblock 2015.
\newblock Assessing the comparative impact of usda frontier and remote area
  (far) codes.
\newblock
  \url{https://nosorh.org/wp-content/uploads/2015/06/Comparative-Impact-of-FAR-and-Super-Rural-Definitions-June-30-2015.pdf}.

\bibitem[\protect\citeauthoryear{Oddou \bgroup et al\mbox.\egroup
  }{1999}]{oddou1999managing}
Oddou, G.~R., and Derr, C.~B.
\newblock 1999.
\newblock {\em Managing internationally: A personal journey}.
\newblock Harcourt College Pub.

\bibitem[\protect\citeauthoryear{Ortega-Santos}{2019}]{ortega2019crowdsourcing}
Ortega-Santos, I.
\newblock 2019.
\newblock Crowdsourcing for hispanic linguistics: Amazon’s mechanical turk as
  a source of spanish data.
\newblock {\em Borealis--An International Journal of Hispanic Linguistics}
  8(1):187--215.

\bibitem[\protect\citeauthoryear{Posch \bgroup et al\mbox.\egroup
  }{2018}]{posch2018characterizing}
Posch, L.; Bleier, A.; Fl{\"o}ck, F.; and Strohmaier, M.
\newblock 2018.
\newblock Characterizing the global crowd workforce: A cross-country comparison
  of crowdworker demographics.
\newblock {\em arXiv preprint arXiv:1812.05948}.

\bibitem[\protect\citeauthoryear{PV}{2007}]{AmazonMe16:online}
PV, S.
\newblock 2007.
\newblock Amazon mechanical turk high on india; introduces rupee payment option
  – gigaom.
\newblock
  \url{https://gigaom.com/2007/05/09/419-amazon-mechanical-turk-high-on-india-introduces-rupee-payment-option}.

\bibitem[\protect\citeauthoryear{Ratcliffe \bgroup et al\mbox.\egroup
  }{2016}]{ratcliffe2016defining}
Ratcliffe, M.; Burd, C.; Holder, K.; and Fields, A.
\newblock 2016.
\newblock Defining rural at the us census bureau.
\newblock {\em American community survey and geography brief} 1:8.

\bibitem[\protect\citeauthoryear{Robertson}{2019}]{nytimes}
Robertson, C.
\newblock 2019.
\newblock They were promised coding jobs in appalachia. now they say it was a
  fraud. - the new york times.
\newblock
  \url{https://www.nytimes.com/2019/05/12/us/mined-minds-west-virginia-coding.html}.

\bibitem[\protect\citeauthoryear{Ross \bgroup et al\mbox.\egroup
  }{2009}]{ross2009turkers}
Ross, J.; Zaldivar, A.; Irani, L.; and Tomlinson, B.
\newblock 2009.
\newblock Who are the turkers? worker demographics in amazon mechanical turk.
\newblock {\em Department of Informatics, University of California, Irvine,
  USA, Tech. Rep}.

\bibitem[\protect\citeauthoryear{Salehi \bgroup et al\mbox.\egroup
  }{2015}]{salehi2015we}
Salehi, N.; Irani, L.~C.; Bernstein, M.~S.; Alkhatib, A.; Ogbe, E.; and
  Milland, K.
\newblock 2015.
\newblock We are dynamo: Overcoming stalling and friction in collective action
  for crowd workers.
\newblock In {\em Proceedings of the 33rd annual ACM conference on human
  factors in computing systems},  1621--1630.

\bibitem[\protect\citeauthoryear{Sandberg \bgroup et al\mbox.\egroup
  }{2020}]{sandberg2020development}
Sandberg, C.; Gray, T.; and Kiran, S.
\newblock 2020.
\newblock Development of a free online interactive naming therapy for bilingual
  aphasia.
\newblock {\em American journal of speech-language pathology}.

\bibitem[\protect\citeauthoryear{Sannon \bgroup et al\mbox.\egroup
  }{2019}]{sannon2019privacy}
Sannon, S., and Cosley, D.
\newblock 2019.
\newblock Privacy, power, and invisible labor on amazon mechanical turk.
\newblock In {\em Proceedings of the 2019 CHI Conference on Human Factors in
  Computing Systems},  1--12.

\bibitem[\protect\citeauthoryear{Sharpe~Wessling \bgroup et al\mbox.\egroup
  }{2017}]{sharpe2017mturk}
Sharpe~Wessling, K.; Huber, J.; and Netzer, O.
\newblock 2017.
\newblock Mturk character misrepresentation: Assessment and solutions.
\newblock {\em Journal of Consumer Research} 44(1):211--230.

\bibitem[\protect\citeauthoryear{Stephany \bgroup et al\mbox.\egroup
  }{2020}]{stephany2020distancing}
Stephany, F.; Dunn, M.; Sawyer, S.; Lehdonvirta, V.; et~al.
\newblock 2020.
\newblock Distancing bonus or downscaling loss? the changing livelihood of us
  online workers in times of covid-19.
\newblock Technical report, Center for Open Science.

\bibitem[\protect\citeauthoryear{Sutherland \bgroup et al\mbox.\egroup
  }{2020}]{sutherland2020work}
Sutherland, W.; Jarrahi, M.~H.; Dunn, M.; and Nelson, S.~B.
\newblock 2020.
\newblock Work precarity and gig literacies in online freelancing.
\newblock {\em Work, Employment and Society} 34(3):457--475.

\bibitem[\protect\citeauthoryear{Suzuki \bgroup et al\mbox.\egroup
  }{2016}]{suzuki2016atelier}
Suzuki, R.; Salehi, N.; Lam, M.~S.; Marroquin, J.~C.; and Bernstein, M.~S.
\newblock 2016.
\newblock Atelier: Repurposing expert crowdsourcing tasks as micro-internships.
\newblock In {\em Proceedings of the 2016 CHI Conference on Human Factors in
  Computing Systems},  2645--2656.

\bibitem[\protect\citeauthoryear{Thiede \bgroup et al\mbox.\egroup
  }{2018}]{thiedeLichter2018working}
Thiede, B.~C.; Lichter, D.~T.; and Slack, T.
\newblock 2018.
\newblock Working, but poor: The good life in rural america?
\newblock {\em Journal of Rural Studies} 59:183--193.

\bibitem[\protect\citeauthoryear{Tietze \bgroup et al\mbox.\egroup
  }{2002}]{tietze2002work}
Tietze, S., and Musson, G.
\newblock 2002.
\newblock When ‘work’meets ‘home’ temporal flexibility as lived
  experience.
\newblock {\em Time \& Society} 11(2-3):315--334.

\bibitem[\protect\citeauthoryear{Van~Hoye \bgroup et al\mbox.\egroup
  }{2016}]{van2016social}
Van~Hoye, G.; Weijters, B.; Lievens, F.; and Stockman, S.
\newblock 2016.
\newblock Social influences in recruitment: When is word-of-mouth most
  effective?
\newblock {\em International Journal of Selection and Assessment} 24(1):42--53.

\bibitem[\protect\citeauthoryear{Vasantha \bgroup et al\mbox.\egroup
  }{2014}]{vasantha2014social}
Vasantha, A.; Vijayumar, G.; Corney, J.; Acur~Bakir, N.; Lynn, A.; Jagadeesan,
  A.~P.; Smith, M.; and Agarwal, A.
\newblock 2014.
\newblock Social implications of crowdsourcing in rural scotland.
\newblock {\em International Journal of Social Science \& Human Behavior Study}
  1(3):47--52.

\bibitem[\protect\citeauthoryear{Williams \bgroup et al\mbox.\egroup
  }{2019}]{williams2019perpetual}
Williams, A.~C.; Mark, G.; Milland, K.; Lank, E.; and Law, E.
\newblock 2019.
\newblock The perpetual work life of crowdworkers: How tooling practices
  increase fragmentation in crowdwork.
\newblock {\em Proceedings of the ACM on Human-Computer Interaction}
  3(CSCW):1--28.

\bibitem[\protect\citeauthoryear{Wood \bgroup et al\mbox.\egroup
  }{2018}]{wood2018workers}
Wood, A.~J.; Lehdonvirta, V.; and Graham, M.
\newblock 2018.
\newblock Workers of the internet unite? online freelancer organisation among
  remote gig economy workers in six asian and african countries.
\newblock {\em New Technology, Work and Employment} 33(2):95--112.

\bibitem[\protect\citeauthoryear{Ye \bgroup et al\mbox.\egroup
  }{2017}]{ye2017does}
Ye, T.; You, S.; and Robert~Jr, L.~P.
\newblock 2017.
\newblock When does more money work? examining the role of perceived fairness
  in pay on the performance quality of crowdworkers.
\newblock In {\em Eleventh International AAAI Conference on Web and Social
  Media}.

\bibitem[\protect\citeauthoryear{Zahnd \bgroup et al\mbox.\egroup
  }{2019}]{zahnd2019challenges}
Zahnd, W.~E.; Askelson, N.; Vanderpool, R.~C.; Stradtman, L.; Edward, J.;
  Farris, P.~E.; Petermann, V.; and Eberth, J.~M.
\newblock 2019.
\newblock Challenges of using nationally representative, population-based
  surveys to assess rural cancer disparities.
\newblock {\em Preventive medicine} 129:105812.

\bibitem[\protect\citeauthoryear{Zyskowski \bgroup et al\mbox.\egroup
  }{2015}]{zyskowski2015accessible}
Zyskowski, K.; Morris, M.~R.; Bigham, J.~P.; Gray, M.~L.; and Kane, S.~K.
\newblock 2015.
\newblock Accessible crowdwork? understanding the value in and challenge of
  microtask employment for people with disabilities.
\newblock In {\em Proceedings of the 18th ACM Conference on Computer Supported
  Cooperative Work \& Social Computing},  1682--1693.

\end{thebibliography}

\end{document}